\documentstyle[epsfig]{aipproc}

\begin{document}
\title{Nuclear Halo and Molecular States}

\author{N.A.~Orr}
\address{Laboratoire de Physique Corpusculaire, \\
IN2P3--CNRS, ISMRa et Universit\'e de Caen, \\
14050 Caen Cedex, France}

\maketitle

\begin{abstract}

Significant advances have been made in recent years in the exploration of clustering
in light nuclei. This progress has arisen not only from the investigation of 
new systems, but also through the development and application of novel 
probes.  This paper will briefly review selected topics concerning halo and molecular states
in light nuclei through examples provided by the neutron-rich Be isotopes.

\end{abstract}

\section*{Introduction}

Clustering within nuclei is a widespread phenomenom which takes on many guises across
the nuclear landscape.  Until relatively recently cluster studies
have been confined to on or near the line of beta stability where the r\^ole
of $\alpha$-clustering has long been 
established \cite{Freerreview}.  
As clustering is expected to manifest itself most
strongly near thresholds \cite{Ikeda}, 
exotic structures might
be expected to form in very neutron (or proton) rich systems.
Over the last decade the exploration of clustering in nuclei far from stability
has become technically feasible as demonstrated most clearly by the discovery and
subsequent probing of the nuclear halo \cite{Han95}. 
Whilst an excess of
neutrons (or protons) may na\"ively be expected to dilute any underlying $\alpha$-cluster
structures, theoretical \cite{Sey81,AMD} and experimental work \cite{vonO,Fre99}
indicate that molecular type structures such as $\alpha$-chains ``bound'' by valence
neutrons may exist.  

In the present paper a number of selected topics concerning the study of halo and
molecular states in light, neutron-rich nuclei will be reviewed.  As they exhibit
many of the facets of clustering and structural evolution far from stability 
the neutron-rich Be isotopes have
been chosen as examples.  In parallel, 
the techniques which have been developed to aid in probing the structure of such nuclei
far from stability will also be discussed.

\section*{Reaction Spectroscopy}

\subsection*{Single-Nucleon Transfer Reactions}

\begin{figure}

\epsfig{file=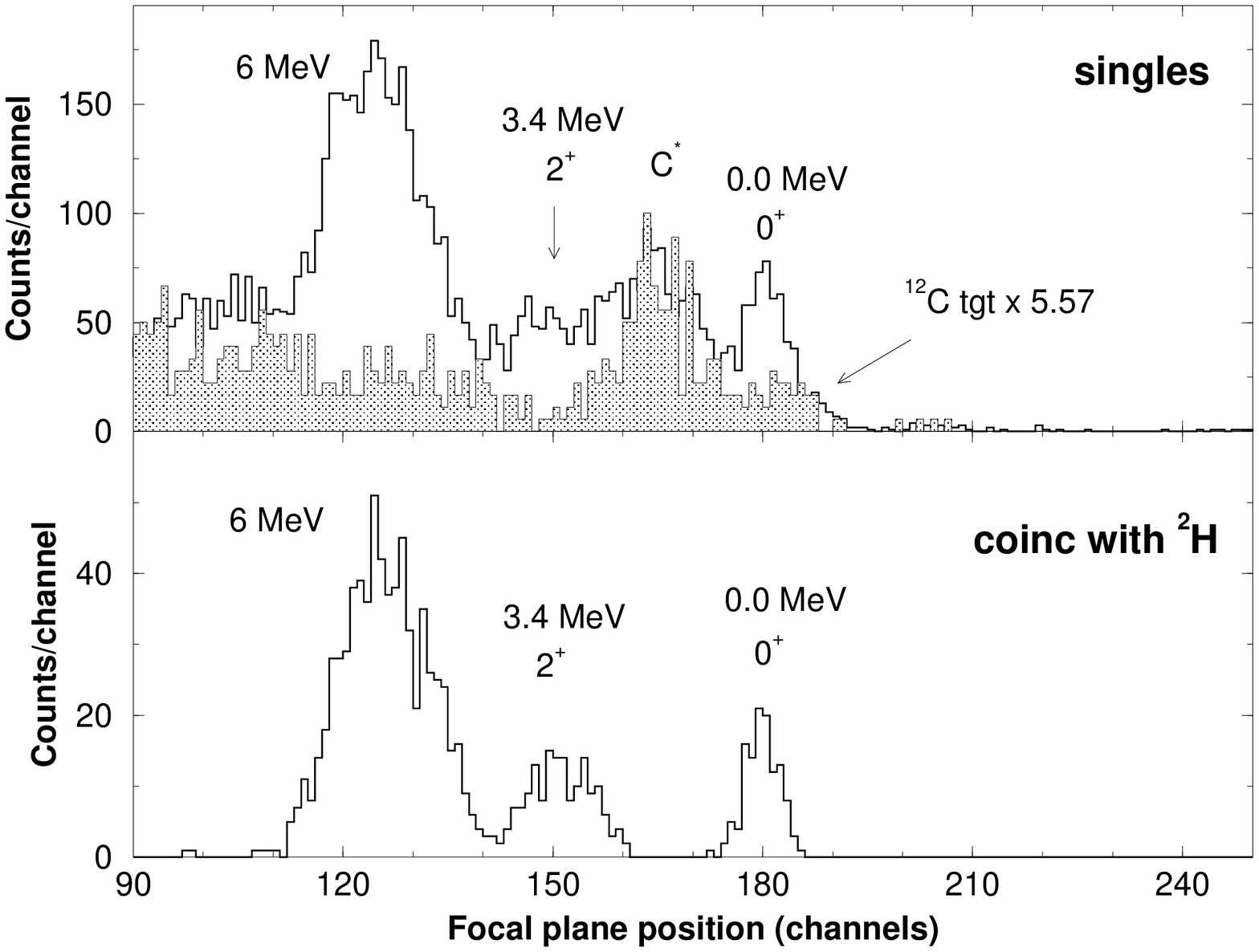,height=6cm}

\vspace*{-8.3cm}

\hspace*{1cm}

\rightline{\epsfig{file=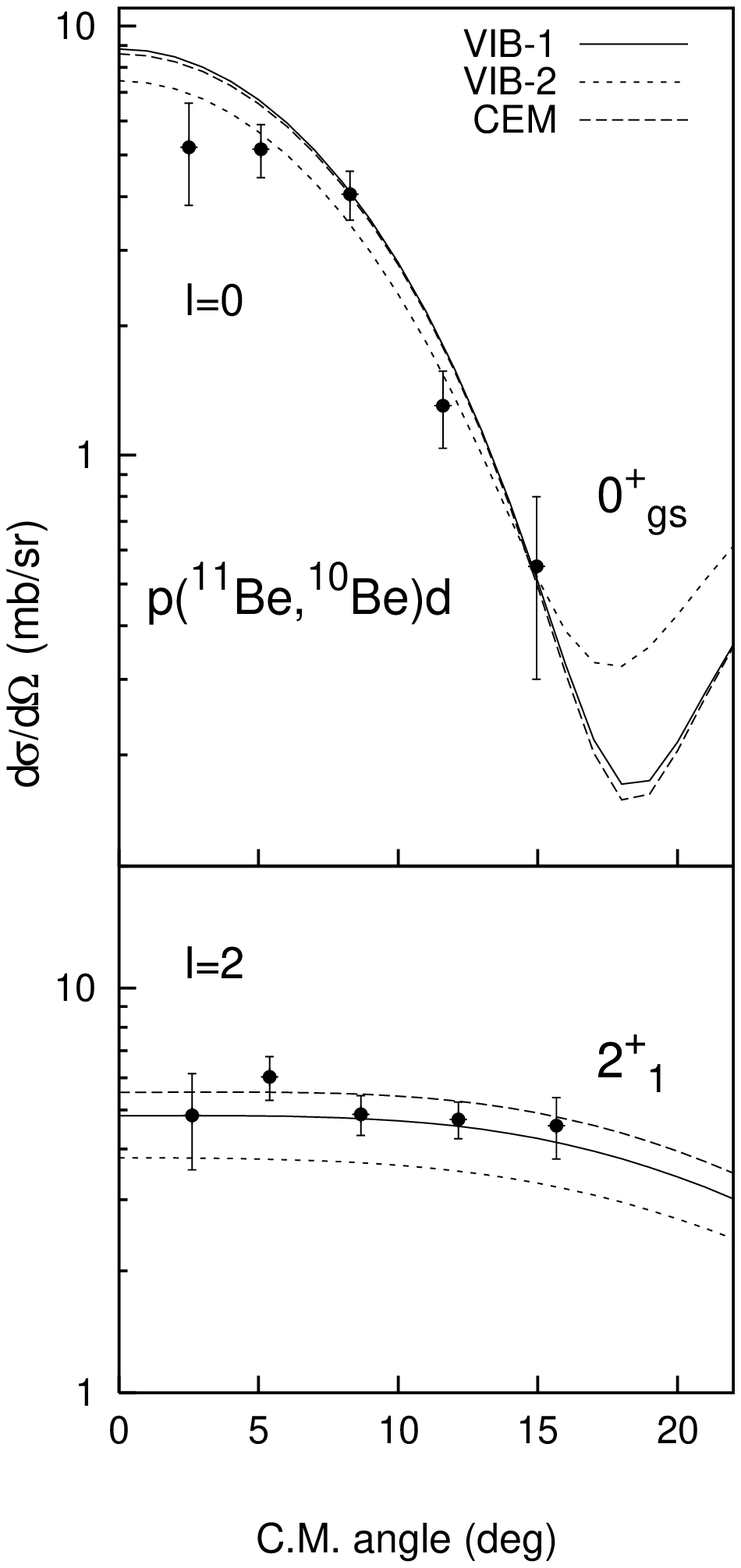,height=10cm}}

\caption{Left: Position spectra for $^{10}$Be ions from the p($^{11}$Be,$^{10}$Be)d 
reaction on a (CH$_2$)$_n$ target.
Right: Angular distributions for transfer to the $^{10}$Be ground and 2$^+$ states.  
Comparison is
made with various core coupling descriptions for $^{11}$Be \protect\cite{For99,Win00}.}

\end{figure}

One of the issues of prime importance in the study of nuclei far from stability is 
the extraction of reliable spectroscopic information.  For nuclei on or near stability
light-ion (p, d, ...) induced single-nucleon transfer reactions have long been one
of the tools of choice.  The application of such reactions to nuclei far from
stability presents a number of experimental challenges \cite{Win97,Cat00}.  Most
significantly, whilst the cross sections are moderately high
($\sim$1-10~mb/sr) the short halflives of the nuclei of interest dictate that
beams of these nuclei be used in inverse kinematics reactions.
The consequent 
constraints on the final excitation energy
and angular resolution limit the target thickness to the order of 1~mg/cm$^2$ and
beams of
at least some 10$^4$~pps are required in order to obtain angular
distributions in a reasonable measurement time.

A recent study \cite{For99,Win00} of the 
structure of the single-neutron halo nucleus 
$^{11}$Be via the p($^{11}$Be,$^{10}$Be)d reaction at 35~MeV/nucleon 
provides a prototypical example of some of the features inherent in such
reaction studies.  Experimentally a magnetic spectrometer
operated in a dispersion matched mode was employed to detect the heavy ejectile
($^{10}$Be).  In order to remove the contamination arising from reactions on
the C in the (CH$_2$)$_n$ target an array of large area silicon detectors
was used to detect the coincident deuterons (figure 1).  With a beam intensity of
$\sim$3$\times$10$^4$~pps angular distributions of reasonable quality were 
acquired in some 72 hours of running (figure 1).  

In addition to the experimental features illustrated by this experiment it should be 
stressed that the analysis
is not model independant.  
For example, not only did the weakly bound nature of $^{11}$Be and the d have to 
be taken into account
but realistic wavefunctions had to be employed -- here coupling to
the strongly deformed $^{10}$Be core needed to be properly accounted for.  
Indeed, the use of the standard separation energy approach to derive the
radial form factors lead to an artifically
high core excited state admixture.  
As detailed in refs \cite{For99,Win00}, careful
analysis indicates that the ground state of $^{11}$Be is dominated (85\%) by the
admixture corresponding to the valence neutron occupying the 2s$_{1/2}$ orbital together
with a modest contribution (15\%) from the core excited (2$^+$$\otimes \nu$1d$_{5/2}$)
configuration.

\subsection*{Single-Nucleon Removal Reactions}

Measurements of one-nucleon removal (or ``knockout'') reactions on light targets
have recently been proposed as a spectroscopic tool for high-energy radioactive beams
\cite{Nav98,Tos99}.  This approach has arisen from the development of reaction
calculations for halo nuclei in which the strong absorption limit \cite{Huf81} and core excited
states are accounted for \cite{Tos99}.  
More specifically, the cross 
sections for the population of a given state of the core fragment ($I^{\pi}_{c}$) 
may be related to spectroscopic factors 
($C^2S(I^{\pi}_{c},nlj)$) 
using an extended version
\cite{Tos99,Tos00} of the spectator-core model \cite{Hus85} to calculate the 
cross section ($\sigma_{sp}$) for the removal of the nucleon ($nlj$). 

\begin{equation}
\sigma(I^{\pi}_{c}) \ = \ \sum_{nlj} C^2S(I^{\pi}_{c},nlj) \sigma_{sp}(nlj,S_n^{eff})
\end{equation}

The corresponding momentum distributions are derived within the same eikonal
formalism
\cite{Sau00} or in a simpler fashion using
the opaque limit of the Serber model \cite{Han96,Esb96}.
As noted above, the integrated cross sections for the population of the 
core excited states 
are directly related to the associated spectroscopic factors.  In analogy with 
transfer reactions, 
the shape of the
core fragment momentum distributions plays the r\^ole of the angular distributions in 
specifying the $l$ of the removed nucleon.
Experimentally the core fragment states are identified by the de-excitation 
$\gamma$-rays emitted in-flight, whilst a high acceptance magnetic spectrograph
is used to determine the momenta.

\begin{figure}
\centerline{\epsfig{file=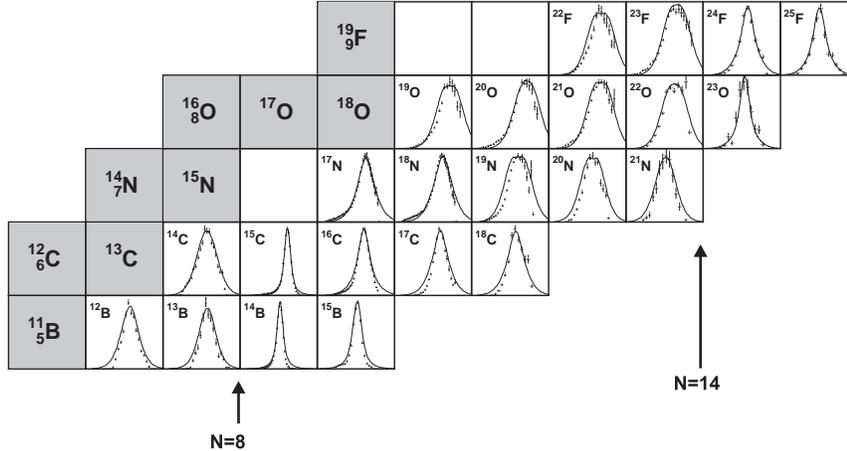,height=6cm}}

\caption{Core fragment longitudinal momentum distributions for one-neutron
removal reactions on C.  The solid lines correspond to Glauber model
calculations \protect\cite{Sau00}.}
\vspace*{-1cm}

\end{figure}

One of the principal virtues offered by high energy nucleon removal 
is the applicability very
far from stability where beam rates are low.  This ability to
function with intensities as low as 1pps is a 
consequence of the large cross sections 
($\sim$10-100~mb), coupled with the high beam energies ($\gtrsim$40~MeV/nucleon)
which allow for the use of thick targets ($\sim$100~mg/cm$^2$).

A particularly clear example\footnote{Studies of other near dripline
and halo nuclei may be found in ref's \cite{Nav98,Aum00,Gui00,Mad00}.} 
of the application of the technique may be found in the measurement of 
single-neutron removal from $^{12}$Be \cite{Nav00}, whereby the only bound core states
are the ground (J$^{\pi}$=1/2$^+$) and 320~keV (1/2$^-$) levels in $^{11}$Be.
In simple terms, the population of the $^{11}$Be ground state provides a measure of
the 2$\hbar\omega$ admixture present in $^{12}$Be.  The results of the experiment,
which indicate only a $\sim$30\% admixture of the $p^2$-configuration in $^{12}$Be, 
provide direct confirmation \cite{She99,Be12beta2} of the breakdown in the N=8 shell closure.

The potential power of the technique is further illustrated by a recent
systematic investigation of a broad range of light, 
neutron-rich psd-shell nuclei \cite{Sau00}.  The inclusive longitudinal momentum 
distributions which were obtained in a single rigidity setting are displayed in 
figure 2 whereby a number of features are immediately     
apparent.  Most notably, the crossing of the N=8 shell and N=14 
sub-shell closures
are associated with a marked reduction in the widths of the core momentum 
distributions (viz, $^{14,15}$B, $^{15,16}$C, $^{23}$O and $^{24,25}$F).  
The former effect arises from the large $\nu$2s$_{1/2}$ admixtures expected 
in the ground states of
the Z=4-6, N=9 isotones \cite{Mad00,Ren97,Alb74,AJZ86}, which also persists for N=10, as
suggested by recent studies of $^{14}$Be \cite{Lab00,Suz99} (see below).
A narrowing of the momentum distributions may also be expected for N=15 and 16,
as in a simple shell model picture the valence neutrons occupy the $\nu$2s$_{1/2}$ 
orbital.  Such results demonstrate that coupled with a high acceptance, broad range
spectrograph, high energy single-nucleon removal reactions offer a powerful means to 
survey structural evolution over a wide range of isospin in a single experiment.

Finally, in the context of high energy reactions, it should also be noted, 
as described by Tostevin and Al-Khalili \cite{Tos00,AlK96}, that
few-body Glauber model analyses of total reaction 
cross sections can also provide important constraints on the ground state 
wavefunctions of halo-like nuclei.  As presented in the contribution to these proceedings 
by Suzuki, such analyses are now becoming more widespread in their application.

\section*{Nuclear Molecular Clusters}

As alluded to in the introduction, the $\alpha$-particle plays an important
r\^ole in the structure of light $\alpha$-conjugate (A=4n) nuclei \cite{Freerreview}.  This
is a direct consequence of the strongly bound character of the $^{4}$He nucleus
and the weakness of the $\alpha$-$\alpha$ interaction, as evidenced by the unbound nature
of $^{8}$Be.  The persistence of cluster structures for systems lying away
from the line of  beta-stability is well illustrated, as will be discussed here, 
by the beryllium isotopes,
for which the $\alpha$-$\alpha$ system may be regarded as the basis.  

From a theoretical point of view, prescriptions such as the Molecular-Orbital Model (MO)
\cite{Sey81} or the Two-Centre Shell Model (TCSM) \cite{TCSM}, in which valence nucleons
are added to the single-particle orbits arising from the two-centre potential, provide a
successful means to describe the properties of these nuclei. Moreover these orbits may be
viewed as the analogues of the $\sigma$ and $\pi$-orbitals associated with the covalent
binding of atomic molecules. The development of fully fledged 
Antisymmeterised Molecular Dynamics calculations (AMD) \cite{AMD} is of particular 
interest as the
N-nucleon system is modelled without any {\em a priori} imposition of an underlying
cluster structure.  Recent calculations, in particular, suggest the existence of two-centred 
structures in the
Be, B and C isotopic chains with valence neutron density 
distributions exhibiting the features of molecular orbitals \cite{AMD}.

From an experimental perspective, von Oertzen \cite{vonO} has compiled systematic
evidence for the existence of dimers in $^{9-11}$Be and $^{9-11}$B.
In the case of $^{9}$Be, for example, 
the presence of a valence neutron results in a bound
(Borromean) system, the ground and excited states of which may be understood
in terms of a three-body $\alpha$:n:$\alpha$ molecular structure.  
In particular,
the rotational bands based on the ground and low-lying states exhibit large
deformations consistent with the associated molecular configurations.

\begin{figure}
\centerline{\epsfig{file=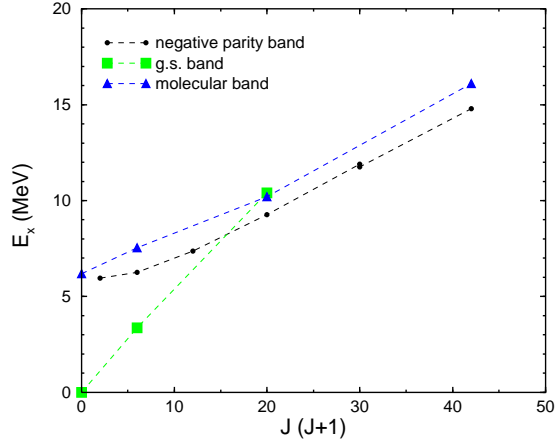,height=6cm}}

\caption{Spin-energy systematics for states observed in $^{10}$Be (from \protect\cite{Fre00}).
The trajectories for the postulated positive and negative parity molecular bands are indicated.}

\end{figure}

In the case of $^{10}$Be, the experimental evidence for molecular configurations
is rather less well documented.  Beyond the 
established 0$^+_2$, 2$^+_2$ and 1$^-_1$ -- 4$^-_1$ states,
the locations of the J=5 \cite{vonO} and 6 members of the of the negative 
parity band, as well as the 
J=4 and 6 members of the positive parity band have been postulated following recent
studies of the $\alpha$-$^{6}$He breakup of $^{10}$Be$^*$ \cite{Soic,Fre00}.
As displayed in figure 3 \cite{Fre00}, the spin-energy trajectories for the bands based on the
0$^+_2$ and 1$^-_1$ states at $\sim$6~MeV are consistent with 
large deformations as expected for molecular-like $\alpha$:2n:$\alpha$
structures.  Moreover, the location of the bandheads just below the threshold for
$\alpha$ + $^{6}$He decay is in accordance with the considerations of Ikeda
describing the formation of clusters \cite{Ikeda}.
Further support for the postulated molecular states may be found in the recent AMD
calculations of Kanada-En'yo and collaborators \cite{AMD}, whereby 
well developed $\alpha$:2n:$\alpha$ configurations are predicted for the 
0$^+_2$ and 1$^-_1$ bands.

Given the existence of such molecular-type structures in $^{10}$Be, the question naturally
arises as to the existance of similar structures even further from stability.
In this context the
dripline nucleus $^{12}$Be has been investigated.  In a recent study\footnote{The existence 
of such structures was hinted at in an earlier study by Korsheninnikov {\em et al.}
\cite{Kor}.} employing the
inelastic excitation of an energetic (35~MeV/nucleon) secondary beam of $^{12}$Be,
evidence has been found in the breakup into $^{6}$He+$^{6}$He of rotational states
(J=4, 6, 8)
in the excitation energy range 10-20~MeV \cite{Fre99}.
As illustrated in figure 4, the inferred momenta of inertia ($\hbar^2/2\Im$=0.15$\pm$0.04~MeV) 
and bandhead 
energy (10.8$\pm$1.8~MeV) of the observed states
are consistent with the cluster decay
of a molecular structure which may be associated with $\alpha$:4n:$\alpha$
configurations.

\begin{figure}

\vspace*{-2.5cm}

\hspace*{+2cm}

\centerline{\epsfig{file=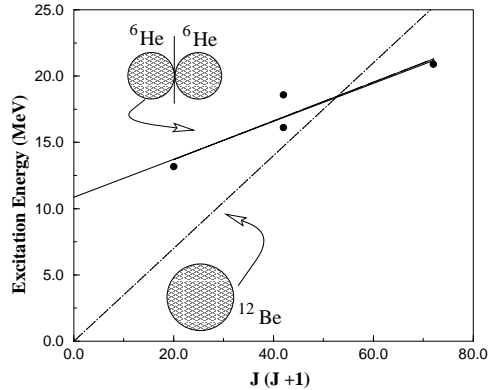,height=11.0cm,angle=270}}

\vspace*{-3.0cm}

\caption{Spin-energy systematics for $^{12}$Be $\rightarrow$ $^6$He + $^6$He (from 
\protect\cite{Fre99}).}

\end{figure}

Further experimental support for the molecular nature of these states 
would be the observation of large spectroscopic factors for the 
associated clusters.  The measurement of partial decay widths represents, however, 
formidable experimental challenges, though the presence of relatively few decay
channels for the states in question may facilitate the measurements.  Of additional
interest is the search for in-band gamma transitions which should also furnish
information on the degree of clustering \cite{Gai}.  Such measurements may be
possible in the near future 
through $\alpha$-pickup reactions -- such as $^{12}$C($^{6,8}$He,$^{10,12}$Be$^*$)$^{8}$Be --
carried out in conjunction with high efficiency charged particle and 
gamma arrays.

\section*{Halo States}

Perhaps the most extreme form of clustering is that exhibited by halo nuclei,
whereby one or more nucleons reside on average well beyond the core potential \cite{Han95}.
In the present discussion the two-neutron halo nucleus $^{14}$Be will be used as an illustrative
example.
In particular, the spectroscopy of $^{13,14}$Be, continuum excitations of $^{14}$Be
and the spatial configuration of the halo neutrons will be addressed.

The tool chosen to investigate $^{14}$Be in the work described here was a 
kinematically complete measurement of the fragments ($^{12}$Be and two neutrons) from
the dissociation of a 35~MeV/nucleon beam of $^{14}$Be on C and Pb targets. 
Such a measurement is now a relatively standard technique and allows
the two-neutron removal cross sections, neutron angular distributions and invariant
mass spectra to be extracted, and the neutron-neutron correlations to
be explored.    

The details of the experiment will not be repeated here as they have already been
described in refs \cite{Lab00,Mar99,Mar00}.  It should be stressed, however,
that one of the principle problems confronting such measurements is the detection
with the highest possible efficiency of two beam velocity neutrons 
at very forward angles with
small relative momenta and minimal cross-talk.  As described in refs \cite{Mar99,Mar00},
the use of a highly granular array arranged in a staggered configuration coupled
with off-line cross-talk rejection algorithms (based on 
kinematic conditions) permit such measurements to be undertaken.

\subsection*{Structure and Continuum Excitations}

\begin{figure}
\centerline{\epsfig{file=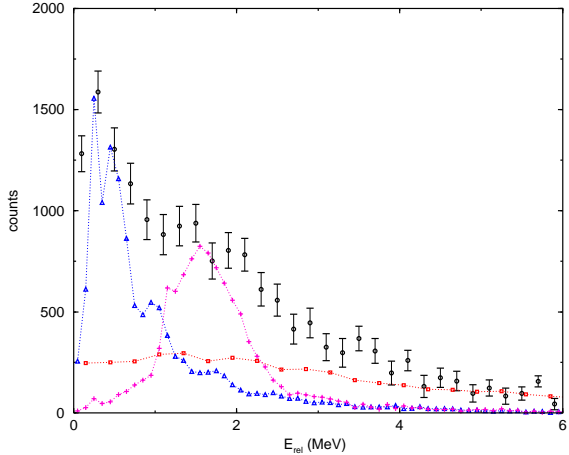,height=7.5cm,angle=270}}

\caption{$^{12}$Be + neutron relative energy spectrum from the dissociation
of $^{14}$Be on C at 35~MeV/nucleon \protect\cite{Jon00}.  The results of a simulation including
states in
$^{13}$Be at 0.5 (triangles) and 2.0~ MeV (crosses), with $\Gamma$=0.5~MeV are also displayed.}
 
\vspace*{-0.3cm}
 
\end{figure}

The results obtained for the two-neutron removal cross sections, $\sigma_{-2n}$, the 
single-neutron angular distributions\footnote{The angular
distributions were well characterised by a Lorentzian lineshape.}, $\Gamma_n$, and 
the associated angle integrated cross sections, $\sigma_n$, are displayed 
in Table 1. 
The average neutron multiplicities ($\overline{m}_n =
\sigma_n/\sigma_{-2n}$), which have also been deduced, are instructive in terms
of the mechanisms leading to dissociation \cite{Bar93}.  For a light target 
the reaction is expected to
proceed via single-neutron removal (absorption or diffraction) followed by the
in-flight decay of $^{13}$Be.  As approximately equal contributions are 
expected for 
absorption and diffraction \cite{Bar93} the average neutron multiplicity should be
1.5, in accordance with that measured. This scenario is also
supported by the single-neutron angular distribution for the C target which is well
reproduced assuming passage via a low-lying resonance in $^{13}$Be
\cite{Lab00,Bar96}. In the case of a heavy target, nuclear and Coulomb dissociation
occur.  Given that Coulomb dissociation results in a
multiplicity of 2, the average multiplicity for dissociation on Pb should be between
1.5 and 2, as observed.


\begin{table}
\begin{tabular}{ccccc}
\tableline
        & $\sigma_{-2n}$ [b] & $\sigma_n$ [b] & $\overline{m}_n$ & $\Gamma_n$ [MeV/c] \\
\tableline
C	& 0.46$\pm$0.04     & 0.75$\pm$0.10  & 1.6$\pm$0.3     & 75$\pm$3 \\
Pb      & 2.3$\pm$0.4       & 4.0$\pm$0.3    & 1.7$\pm$0.2     & 77$\pm$4 \\
Pb(EMD) & 1.45$\pm$0.40     & 2.7$\pm$0.4    & 1.9$\pm$0.6     & 87$\pm$6 \\
\tableline
\end{tabular}
\caption{Measured cross sections, average neutron multiplicities and 
neutron distribution 
momentum widths for the dissociation of $^{14}$Be 
at 35~MeV/nucleon \protect\cite{Lab00}.}
\end{table}


The possible existence of a low-lying state in $^{13}$Be is of considerable importance
in modelling the structure of $^{14}$Be \cite{Tho96}; in particular in fixing the  
$^{12}$Be-n interaction.  Beyond the single-neutron angular distributions, the relative
energy for $^{12}$Be + n events may be reconstructed from the measured momenta.  
The preliminary results of such an analysis \cite{Jon00} are displayed in
figure 5 together with the results of Monte Carlo simulations (see below) in which levels
at 0.5 and 2.0~MeV above threshold are assumed with widths of $\Gamma$=0.5~MeV.  The 
level at 2.0~MeV is that observed in multi-nucleon transfer reaction studies 
\cite{Ost92,Bel},
whilst the former is consistent with a preliminary result reported by the MSU group
\cite{Tho95}.  Given, as discussed above, the appearance of ground states 
dominated by a valence $\nu$2s$_{1/2}$
configuration for the neighbouring N=9 isotones $^{14}$B and $^{15}$C \cite{Sau00,Gui00,Mad00},
together with the predicted continuation of such an inversion for $^{13}$Be \cite{Ren97},
it appears that the ground state is most probably J$^{\pi}$=1/2$^+$.  Analysis of 
the $^{12}$Be - n 
angular correlations should shed further light on this conjecture.

The enhanced cross section for dissociation on the Pb target 
(Table 1) is indicative of a large
EMD contribution. Assuming that the nuclear--Coulomb interference is small, the C
target data may be scaled
to estimate the nuclear contribution to breakup on Pb \cite{Lab00}.  Given a
root-mean-square  radius of 3.2 fm for $^{14}$Be \cite{Suz99}, 
$\sigma_{-2n}^{nucl}(Pb)$ = 0.85$\pm$0.07~b and, consequently, 
$\sigma_{-2n}^{EMD}(Pb)$ = 1.45$\pm$0.40~b. 

The invariant mass spectra, reconstructed from the measured momenta of the beam and
fragments ($^{12}$Be and two neutrons) from breakup, are displayed in figure 6a and
b for the C and Pb targets.  The EMD spectrum (figure 6c) has been deduced 
following subtraction of the estimated  nuclear contribution to 
reactions on Pb and exhibits enhanced strength around 1.5~MeV decay energy ($E_{decay}$). Given
the complex nature of the response function of the present setup, a detailed Monte
Carlo simulation, including the influence of all nonactive materials, was developed
based on the GEANT package \cite{Lab00}.
The results shown in figure 6 were obtained following the descriptions for
dissociation on C and Pb outlined above and after filtering through the
simulation.  In the case of the nuclear induced reactions only the lowest low-lying state in
$^{13}$Be ($E_0 = 0.5$~MeV, $\Gamma_0 = 0.5$~MeV) was assumed to be populated
following the diffraction of one of the halo neutrons. The EMD was
simulated under the assumption that the energy sharing between the $^{12}$Be and the
two neutrons was governed by 3-body phase space.  As shown in figure 6c, the observed
EMD decay energy spectrum could be reproduced assuming a Breit-Wigner lineshape with
$E_0 = 1.8\pm0.1$~MeV and $\Gamma_0 = 0.8\pm0.4$~MeV.  Furthermore,
the corresponding simulations of the single-neutron angular distributions were in
good agreement with those observed \cite{Lab00}.

Thompson and Zhukov have examined  $^{14}$Be within the framework of a 
3-body model in which the $^{12}$Be core was treated as inert\footnote{An inert core 
precludes, ab initio, the existence of
any simple negative parity resonances.} \cite{Tho96} and a number
of trial wavefunctions developed.  Based on the binding energy and matter radius
of $^{14}$Be, together with the known d-wave resonance at 2.0~MeV in
$^{13}$Be, two $^{14}$Be wavefunctions were favoured (both of which required
an s-wave state near threshold in $^{13}$Be as suggested above): the so-called D4  
wavefunction -- 86\% $\nu (2s_{1/2})^{2}$ and
10\% $\nu (1d_{5/2})^{2}$; and C7 -- 29\% $\nu (2s_{1/2})^{2}$ and 67\% $\nu
(1d_{5/2})^{2}$.  The EMD decay energy spectra calculated from the  
corresponding E1
strength functions \cite{Tho96} are compared in
figure 6 with that of the empirical Breit-Wigner lineshape deduced from the
measurements.  The corresponding integrated two-neutron removal cross sections are
1.05~b (D4) and 0.395~b (C7) \cite{Tho96}, compared to the measured value of
1.45$\pm$0.40~b.  Although the strength is predicted to be concentrated at a somewhat
lower energy than that observed, a large $\nu (2s_{1/2})^{2}$ admixture to the valence
neutrons wavefunction is favoured. This conclusion is also supported by the total reaction
cross section measurement of Suzuki {\em et al.} \cite{Suz99} reported in
these proceedings.

A microscopic cluster model has also been used to explore $^{13,14}$Be \cite{Des95}.  
In the case of $^{13}$Be an s-wave
state is predicted very close  to threshold, whilst the energy of the d-wave
resonance is well reproduced.   Significantly, a strong E1 transition
[B(E1)$\approx$1.2e$^2$fm$^2$] centred at $E_{decay}=1.5$~MeV is predicted in
$^{14}$Be,  very close to the structure observed experimentally.  Analysis of the
corresponding energy surface suggests, however, that this transition is not
associated with a true resonance \cite{Des95}.  Consequently, the enhanced strength 
observed near threshold in the EMD invariant mass spectrum may be qualified as a
nonresonant ``soft'' dipole excitation.

In
light of the breakdown in the neutron p-shell closure in $^{12}$Be discussed previously, 
the importance of p-sd cross shell excitations in the $^{14}$Be ground state should be
explored.  In a first instance, the search for core excited states ($^{12}$Be$^*$) in the
dissociation of $^{14}$Be may provide indications for the presence of
such configurations, whilst theoretically the inclusion of a 
deformed core \cite{Be12beta2},
in three-body calculations should be undertaken.

\subsection*{Neutron--Neutron Correlations}

One of the most intriguing questions regarding the description of two-neutron halo systems
is the degree of 
correlation between the neutrons. 
Considering only the intrinsic momentum distributions, a spatially compact dineutron 
will be characterised by $p_{core}=2p_n$, whereas for a 
system with no correlations $p_{core}=\sqrt{2}p_n$.  Unfortunately no unperturbed 
experimental measure of the momentum 
distributions is accessible.
 
\begin{figure}

\epsfig{file=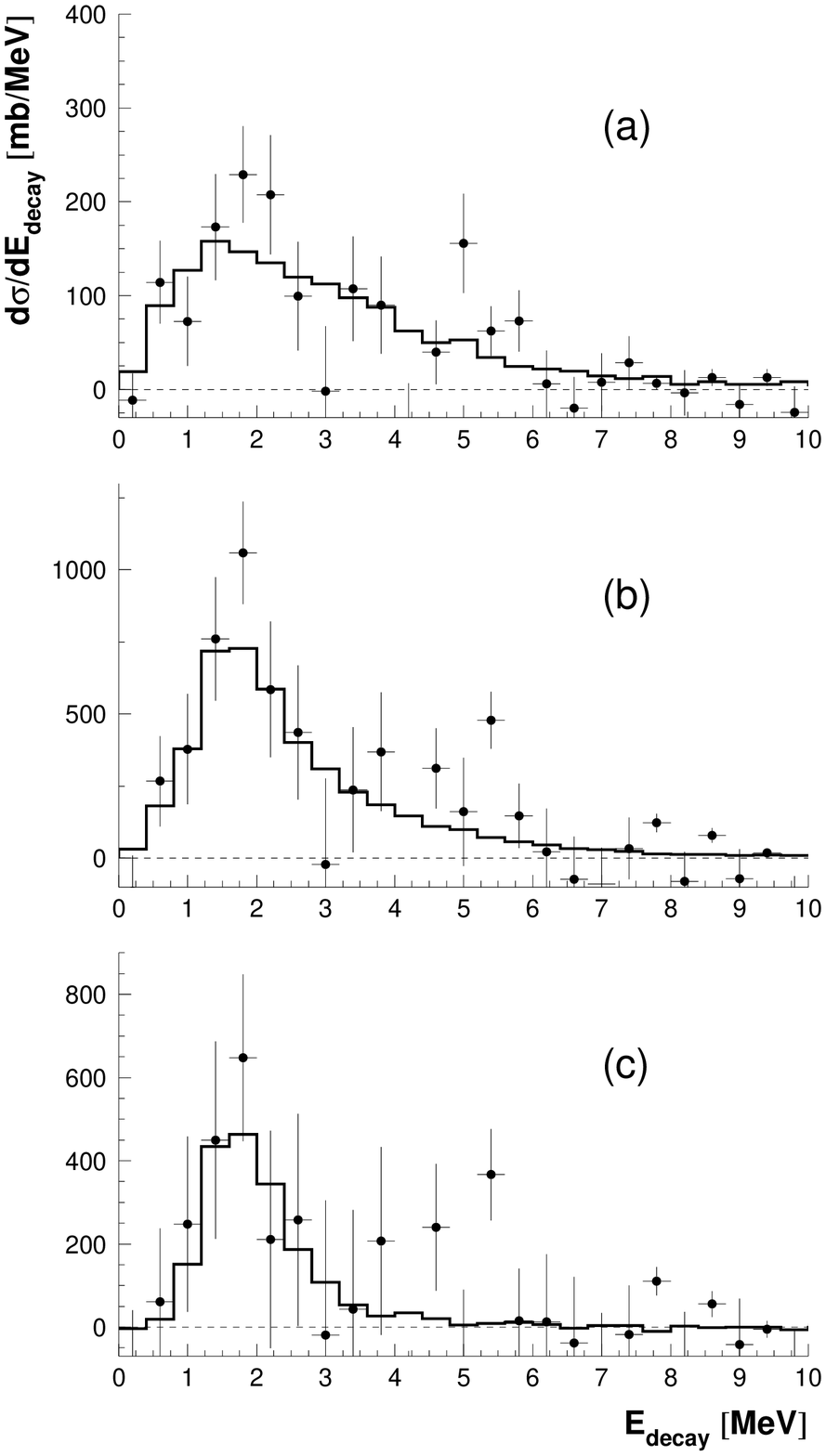,height=9cm}

\vspace*{-10cm}

\rightline{\epsfig{file=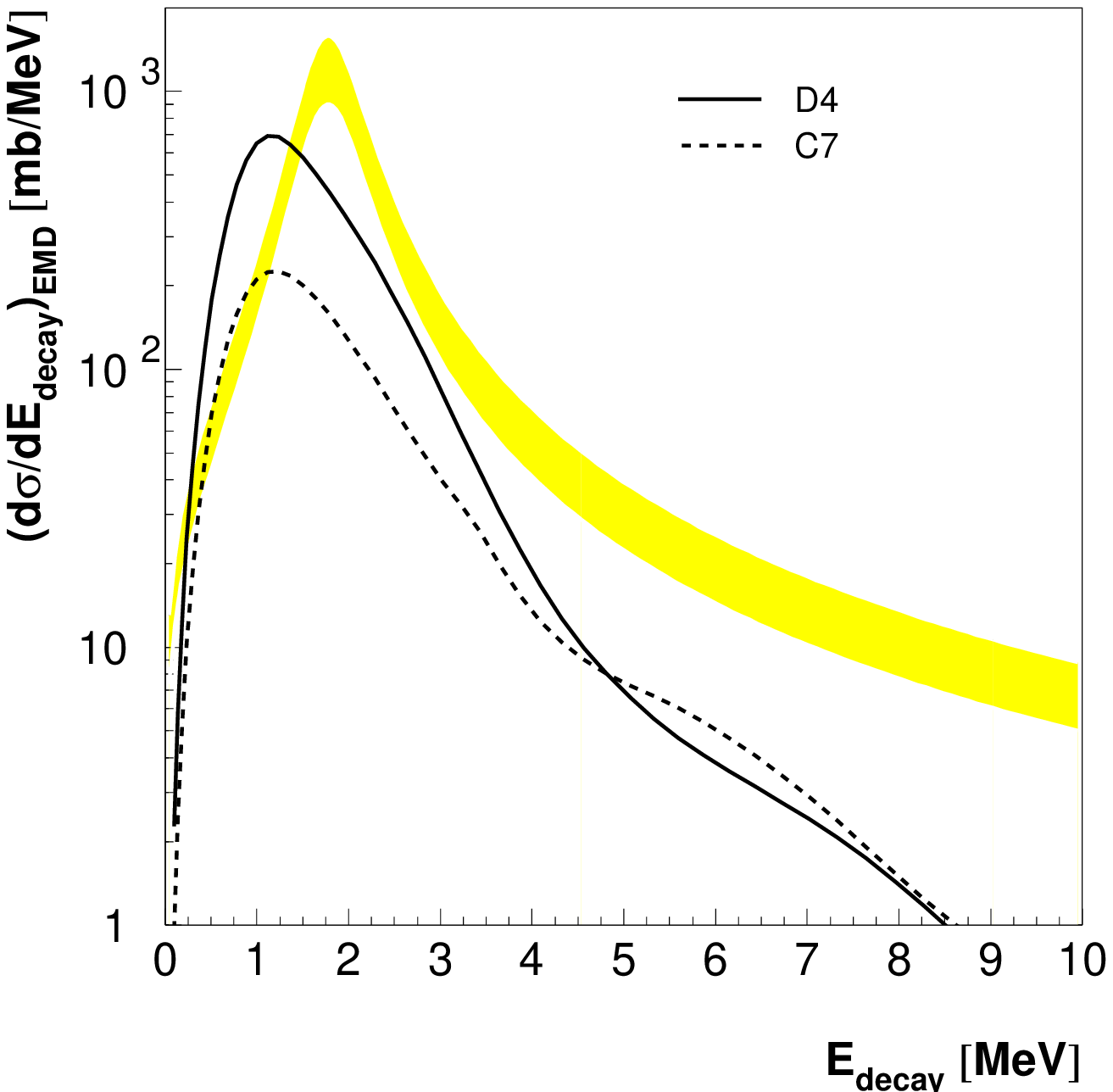,height=9cm}}

\vspace*{0.7cm}

\caption{Left: Reconstructed decay energy spectra for the dissociation of $^{14}$Be on (a) C,
(b) Pb and (c) deduced for EMD on Pb.  Right:  Comparison of the EMD
decay energy spectrum deduced from the measurement (shaded region) with 
3-body calculations (see text).}

\vspace*{-0.3cm}

\end{figure}

Perhaps the method the most well adapted to probing spatial correlations 
is intensity 
interferometry \cite{Boa}.
In this approach the relative motion of the two outgoing neutrons is 
governed by 
the neutron--neutron final-state interaction and quantum statistics 
(Fermi-Dirac), both of which are related to the 
spatial  
characteristics of the source.  Providing that any 
other effects on 
the neutron momenta may be neglected or eliminated in the construction of the 
correlation function, interferometry provides a means to probe the
spatial correlations of the halo neutrons \cite{Mar99}.  

Importantly, owing to
the low momentum content of the halo neutrons, the standard approach to
constructing correlation functions (applied, for example, 
in the measurement of Ieki {\em et al.} \cite{Sac93}) is no longer valid 
\cite{Mar00}.  
In particular, the narrow momentum distributions result in strong 
residual correlations and consequently a significant 
overestimate of the source size.  As 
a result a new iterative technique has recently been developed and applied to
the dissociation on a Pb target of beams of
$^{6}$He, $^{11}$Li and $^{14}$Be \cite{Mar99}.  
Assuming 
that the timescale
between the emission of the two neutrons was short ($\tau_{nn}\lesssim$100~fm/c), as is
most probably the case for EMD, neutron-neutron RMS separations of
5.9$\pm$1.2 ($^{6}$He), 6.6$\pm$1.5 ($^{11}$Li), and 5.4$\pm$1.0~fm ($^{14}$Be)
have been deduced.
These results appear to exclude the presence of any significant
dineutron configurations.  By way of comparison it is interesting to note that
the RMS proton-neutron distance in the deuteron is some 3.8~fm.
Future high statistics experiments emplying a well modelled system such as $^{6}$He 
should allow the emission timescale and source 
sizes to be extracted simultaneously from the longitudinal and transverse
neutron-neutron relative momenta as well as coherent analyses of the neutron-neutron and 
core-neutron correlations.

\section*{Conclusions}

In this paper some of the facets of nuclear halo and molecular states
in light nuclei have been reviewed.  In particular, various illustrative
examples provided by the neutron-rich beryllium isotopes have been discussed.  
In addition, a number of related experimental probes have also been presented.  
Clearly no single technique can furnish a 
complete description of these nuclei and it will require the application of 
a broad range of theoretical and experimental tools  
to obtain a better understanding of the phenomena outlined here.

It is a pleasure to thank my colleagues in the Groupe Exotiques at LPC and in the 
DEMON, CHARISSA and E264/281/295 collaborations who have contributed to the 
work described here.

\end{document}